\documentclass[12pt]{spieman}  
\usepackage{amsmath,amsfonts,amssymb}
\usepackage{graphicx}
\usepackage{setspace}
\usepackage{tocloft}
\usepackage[shortlabels]{enumitem}
\usepackage{array,multirow,tabularx,siunitx}
\usepackage[table,dvipsnames]{xcolor}
\usepackage{adjustbox}
\usepackage{multicol,tcolorbox,hhline}
\usepackage{arydshln}
\usepackage{booktabs}
\usepackage{rotating}
\usepackage{boldline}
\usepackage{amsbsy}
\usepackage[numbers,sort&compress,super]{natbib} 
\usepackage{soul}

\newcolumntype{C}{>{\centering\arraybackslash}p{2.5cm}}

\title{Milky Way and Nearby Galaxies Science with the Single Aperture Large Telescope for Universe Studies (\textit{SALTUS}) Space Observatory}

\author[a,\dag,*]{Rebecca C. Levy}
\author[b,c]{Alexander Tielens}
\author[d]{Justin Spilker}
\author[a]{Daniel P. Marrone}
\author[e,f]{Desika Narayanan}
\author[a]{Christopher K. Walker}
\affil[a]{Department of Astronomy and Steward Observatory, University of Arizona, Tucson, AZ, 85721, USA}
\affil[b]{Leiden Observatory, P.O. Box 9513, Leiden, The Netherlands}
\affil[c]{Astronomy Department, University of Maryland, College Park, MD 20742, USA}
\affil[d]{Department of Physics and Astronomy and George P. and Cynthia Woods Mitchell Institute for Fundamental Physics and Astronomy, Texas A\&M University, 4242 TAMU, College Station, TX 77843-4242, USA}
\affil[e]{Department of Astronomy, University of Florida, 211 Bryant Space Sciences Center, Gainesville, FL 32611 USA}
\affil[f]{Cosmic Dawn Center at the Niels Bohr Institute, University of Copenhagen and DTU-Space, Technical University of Denmark}

\cftpagenumbersoff{figure}
\cftpagenumbersoff{table} 

\begin{document} 
\maketitle

\newcommand{\arcsec}{$^{\prime\prime}$}
\newcommand{\arcmin}{$^{\prime}$}
\newcommand{\saltus}{SALTUS}
\newcommand{\micron}{$\mu$m}
\newcommand{\kms}{km~s$^{-1}$}
\newcommand{\HII}{H{\small II}}
\newcommand{\CII}{[C{\small II}]}
\newcommand{\SIII}{[S{\small III}]}
\newcommand{\SiII}{[Si{\small II}]}
\newcommand{\OI}{[O{\small I}]}
\newcommand{\OIII}{[O{\small III}]}
\newcommand{\NII}{[N{\small II}]}
\definecolor{darkpink}{rgb}{0.922,0.102,0.349}
\newcommand{\change}{}

\newcommand{\apj}{Astrophys. J.}
\newcommand{\apjs}{Astrophys. J. supplements}
\newcommand{\apjl}{Astrophys. J. letters}
\newcommand{\mnras}{Mon. Not. R. Astron. Soc.}
\newcommand{\nat}{Nature}
\newcommand{\aj}{Astronom. J.}
\newcommand{\aap}{Astron. Astrophys.}
\newcommand{\pasp}{PASP}
\newcommand{\araa}{Ann. Rev. Astron. Astrophys.}
\newcommand{\jatis}{JATIS}

\begin{abstract}
This paper presents an overview of the Milky Way and nearby galaxies science case for the \textit{Single Aperture Large Telescope for Universe Studies} (SALTUS) far-infrared NASA probe-class mission concept. SALTUS offers enormous gains in spatial resolution and spectral sensitivity over previous far-IR missions, thanks to its cold ($<$40~K) 14-m primary mirror. Key Milky Way and nearby galaxies science goals for SALTUS focus on understanding the role of star formation in feedback in the Local Universe. In addition to this science case, SALTUS would open a new window to to of Galactic and extragalactic communities in the 2030s, enable fundamentally new questions to be answered, and be a far-IR analog to the near- and mid-IR capabilities of JWST. This paper summarizes the Milky Way and nearby galaxies science case and plans for notional observing programs in both guaranteed and guest (open) time.
\end{abstract}


{\noindent \footnotesize\textbf{\dag}National Science Foundation Astronomy \& Astrophysics Postdoctoral Fellow}

{\noindent \footnotesize\textbf{*}\linkable{rebeccalevy@arizona.edu} }

\begin{spacing}{1.5}   

\section{Introduction}
\label{sec:intro}

The Single Aperture Large Telescope for Universe Studies (\saltus) observatory is a far-infrared (FIR) mission concept proposed to NASA's recent APEX call for proposals \citep{Chin2024JATIS}. Its unique 14-m primary mirror and sensitive FIR ($\approx30-700$\,\micron) instrumentation would provide orders-of-magnitude improvements over past and other proposed FIR space missions. The very large primary mirror size results unprecedented $\sim$1\arcsec\ spatial resolution in the FIR, making \saltus\ a well-matched bridge in wavelength between the James Webb Space Telescope (JWST) and the Atacama Large Millimeter/submillimeter Array (ALMA). The overall telescope, receivers, and spacecraft architectures are described in detail in a series of papers elsewhere in this issue\citep{Arenberg2024JATIS,Kim2024JATIS,Harding2024JATIS}. Briefly, the two instruments planned for \saltus\ are SAFARI-Lite and HiRX. SAFARI-Lite is a direct detection spectrometer providing medium resolution ($R\sim300$) spectroscopy over the entire $34-230$\,\micron\ wavelength range simultaneously\citep{Roelfsema2024JATIS}. HiRX is a multi-pixel, multi-band heterodyne receiver ($R\sim10^5-10^7$), enabling sub-\kms\ spectroscopy of key spectral lines for interstellar medium (ISM) science\citep{daSilva2024JATIS}.

This paper provides an overview of the Milky Way and nearby galaxies science uniquely enabled by \saltus. Accompanying papers in this issue describe the capabilities of \saltus\ for high-redshift galaxies\citep{Spilker2024JATIS}, star and planet formation\citep{Schwarz2024JATIS}, and solar system observations\citep{Anderson2024JATIS}. In the remainder of this section, we provide background on the key principles and tracers of processes in the ISM of galaxies and the Milky Way. Section \ref{sec:performance} highlights the \saltus\ performance characteristics most relevant for Milky Way and nearby galaxies science. In Section \ref{sec:keygoals}, we present several key measurements, enabled by \saltus, that answer key open questions in the fields of Milky Way and nearby galaxies science.

\subsection{\change{Key Cosmic Ecosystems Questions}}
\label{sec:decadal}

\change{The 2020 Astronomy Decadal Survey\citep{astro2020} has laid out a number of key science questions pertaining to our \textit{Cosmic Ecosystem} that drive the science priorities for the next decade and beyond. While the capabilities of SALTUS would enable substantial progress towards the majority of the \textit{Cosmic Ecosystems} questions, here we focus on the following key decadal questions (DQs) as they pertain to observing the ISM of the MW and nearby galaxies.}
\begin{enumerate}
    \itemsep0em
    \item \change{How do star-forming structures arise from, and interact with, the diffuse interstellar medium?}
        \begin{enumerate}\itemsep0em
            \item \change{How does injection of energy, momentum, and metals from stars (``stellar feedback”) drive the circulation of matter between phases of the ISM and circumgalactic medium?}
        \end{enumerate}
    \item \change{What regulates the structures and motions within molecular clouds?}
        \begin{enumerate}\itemsep0em
            \item \change{What is the origin and prevalence of high-density structures in molecular clouds, and what role do they play in star formation?}
            \item \change{What generates the observed chemical complexity of molecular gas?}
        \end{enumerate}
    \item \change{How do gas, metals, and dust flow into, through, and out of galaxies?}
        \begin{enumerate} \itemsep0em
            \item \change{The production, distribution, and cycling of metals.}
            \item \change{The coupling of small-scale energetic feedback processes to the larger gaseous reservoir.}
        \end{enumerate}
    \item \change{How do the histories of galaxies and their dark matter halos shape their observable properties?}
        \begin{enumerate}
            \item \change{Connecting local galaxies to high-redshift galaxies.}
            \item \change{The evolution of morphologies, gas content, kinematics, and chemical properties of galaxies.}
        \end{enumerate}

\end{enumerate}

\subsection{Stellar Feedback and the Ecology of Galaxies}
\label{sec:feedback}

\begin{figure*}[t]
\centering{\includegraphics[width=0.75\textwidth]{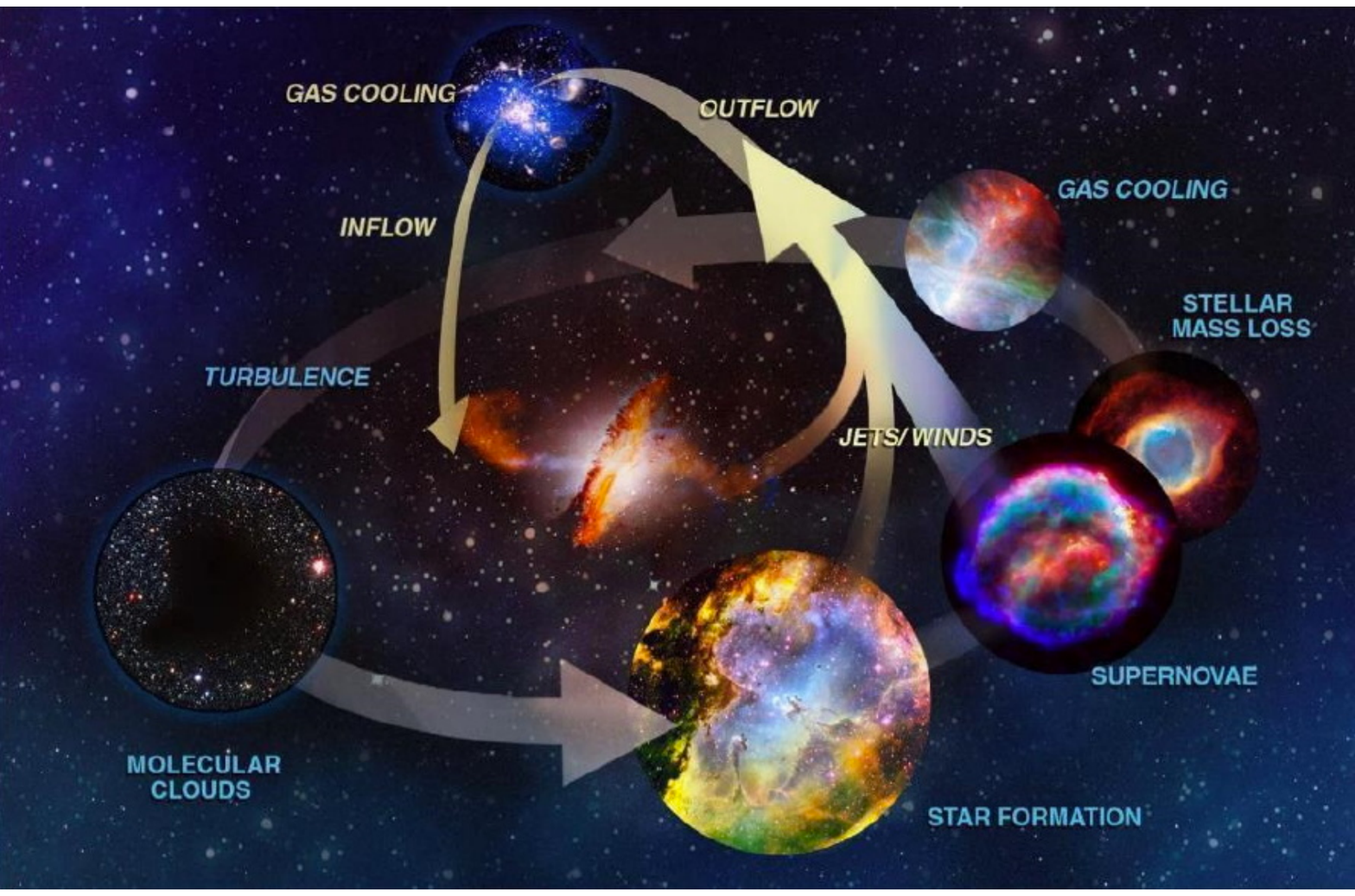}}
\caption{The role of feedback and the lifecycle of galaxies. Stars form inside dense cloud. The mechanical and radiative energy injected during their evolution from the protostellar to the main sequence to the supernova phase will heat interstellar gas, inject turbulence into the medium, disrupt molecular clouds, and create the various phases of the ISM. During the asymptotic giant branch phase and the supernova phase, newly synthesized elements will be ejected, slowly enriching the ISM. The concerted action of massive stars in rich OB associations will drive the formation of chimneys that vent enriched gas into the lower halo and even the intergalactic medium. Halo gas will condense in clouds that rain back onto the galactic plane. The concerted action of these processes drives the evolution of galaxies. Figure taken from \citep{astro2020}. \change{SALTUS will enable key measurements across the this feedback and life cycle at unprecedented spatial and spectral resolutions.}} 
\label{fig:lifecycle}
\end{figure*} 

Stellar feedback plays a central role in galactic ecology (Figure~\ref{fig:lifecycle}) and is a key question posted in the 2020 Astronomy Decadal Survey (e.g., \change{DQs 1 and 3}). During their formation phase, protostellar jets and winds stir up their environment and greatly modify the structure of the clouds in which they form and hence the star formation process.  During their main sequence phase, the interaction of massive stars with their environments regulates the evolution of galaxies \change{(DQs 1a and 3b)}. Mechanical and radiative energy input by massive stars stir up and heat the gas and control cloud and intercloud phases of the ISM. The dominant mode of stellar feedback changes with the stellar/cluster mass and as the star/cluster evolves\citep{Krumholz2019,Olivier2021,Barnes2021}. Stellar feedback also governs the star formation efficiency of molecular clouds\citep{Elmegreen2011,Hopkins2014,Kennicutt2012,Krumholz2019}. On the one hand, stellar feedback can lead to a shredding of the nascent molecular cloud within a few cloud free-fall times thereby halting star formation\citep{Matzner2002,Geen2016,Kim2018}. On the other hand, massive stars may also provide positive feedback to star formation as gravity can more easily overwhelm cloud-supporting forces in swept-up compressed shells\citep{Elmegreen1977,Zavagno2010}. The Spitzer/GLIMPSE galactic plane survey has revealed hundreds of parsec scale bubbles associated with O and B stars, attesting to the importance of stellar feedback in stirring up the ISM\citep{Churchwell2006}. \change{The expansion of these bubbles may be driven by thermal pressure of the warm ionized gas, radiation pressure by stellar photons, or the mechanical energy input by stellar winds}\citep{tielens2005,PortegiesZwart2010,Olivier2021}. The PHANGS survey has revealed larger (up to $\approx$1~kpc) bubbles in nearby galaxies, likely driven by mechanical energy from stellar winds and (clustered) supernovae\citep{Watkins2023,Barnes2023}. As massive star formation dominates the energetics and feedback in star forming galaxies, properly accounting for the star formation feedback is a critical ingredient of galaxy evolution models and validating the subgrid feedback descriptions in these hydrodynamic studies is of paramount importance \change{\citep{Kim2017,Grudic2021,Grudic2023,Lancaster2024}.}

\subsection{Tracers of the ISM}
\label{sec:tracers}

Much of the interaction of massive stars with their environment occurs in regions shrouded in dust and the far-infrared -- where dust extinction is minimal -- provides many key diagnostic transitions that measure the physical conditions in these regions and trace their evolution. Specifically, this spectral range hosts atomic and ionic fine-structure transitions, covering
a wide range in critical density ($100-500$ K; $3\times 10^2-3\times 10^4$ cm$^{-3}$)\citep{tielens2005}.

\change{During the protostellar and main sequence phases, mechanical energy input by stars drives strong shock waves into the surrounding gas, sweeping it up in dense shells. The far-infrared provides a convenient probe of the temperature structure through the slew of CO pure rotational transitions accessible with SAFARI-Lite and HiRX \citep{tielens2021}. The chemical processing of the gas by the shock starts with the sputtering of ice mantles dominated by H$_2$O and to a lesser extent NH$_3$. This is the first step in the build up of chemical complexity in regions of star formation \change{(DQs 2b, 3a, 4b)}.}

The physical conditions in photodissociation regions (PDRs) can be studied through the [OI] 63 \&\ 145 $\mu$m, [CII] 158 $\mu$m, [SiII] 35 $\mu$m fine-structure transitions, as well as high-$J$ CO rotational transitions \change{(e.g., $J=10-9$ at 260~\micron\ through 
$J=19-18$ at 137~\micron)}, in addition to $^{13}$CO isotopes and the deuterated hydrogen (HD) molecule. \change{The high sensitivity will bring many higher order CO transitions into SALTUS purview, as well.} All of these transitions are known to be bright in PDRs\citep{tielens1985,hollenbach1997,joblin2018}. Together, these transitions span a wide range in critical densities ($\simeq 3\times 10^3-10^6$~cm$^{-3}$) and excitation energies ($\simeq 100-500$~K) and thus provide powerful tools to determine physical conditions in these neutral gas regions. The strong far-UV irradiation of PDRs also leaves it imprint on the chemical composition, leading to a strong stratification of the molecular composition ranging from a surface layer characterized by small hydride radicals to the deeper zones dominated by heavier species, including CO, HCN, etc.\citep{sternberg1989}. \change{The first steps toward molecular complexity start with the formation of simple hydride radicals (DQ 2b).} Because of their large moments of inertia, light hydride radicals have their main transitions in the far-IR. Herschel has detected strong emission from HD, OH, H$_2$O, H$_2$O$^+$, H$_3$O$^+$, CH, CJ$^+$, SH, HF, and H$_2$Cl$^+$ in the Orion Bar \citep{nagy2017}. 

\change{Finally, the [NeIII] 36 $\mu$m, [SIII] 33.4 $\mu$m, [NII] 122 \&\ 205 $\mu$m, [NIII] 57 $\mu$m, and [OIII] 52 \&\ 88 $\mu$m lines trace ionized gas in HII regions around hot young stars, and provide a measure of the gas density and the hardness of the stellar radiation fields \citep{leticia2002}. Importantly, these transitions are not temperature sensitive. Therefore, galactic elemental abundance gradients (DQ 4b) can be accurately assessed by combining measurements of these species in the far-IR with mid-IR neutral hydrogen recombination lines provided by JWST.}

\section{SALTUS Performance for Milky Way and Nearby Galaxies Science}
\label{sec:performance}

\change{While the \saltus\ telescope, instrumentation, and spacecraft design principles are described in detail in accompanying papers in this issue\citep{Chin2024JATIS,Arenberg2024JATIS,Kim2024JATIS,Harding2024JATIS,Roelfsema2024JATIS,daSilva2024JATIS}, here we briefly describe the performance metrics most relevant for the Milky Way and nearby galaxies science community.} We envision that the SAFARI-Lite grating spectrometer will be the workhorse instrument for this community. Additionally, key spectral lines (e.g., water, \CII, etc) will be observable with the HiRX heterodyne instrument at very high spectral resolution. 

\subsection{Wavelength Coverage}
\label{ssec:wavelength}

\begin{figure*}
\centering{\includegraphics[width=\textwidth]{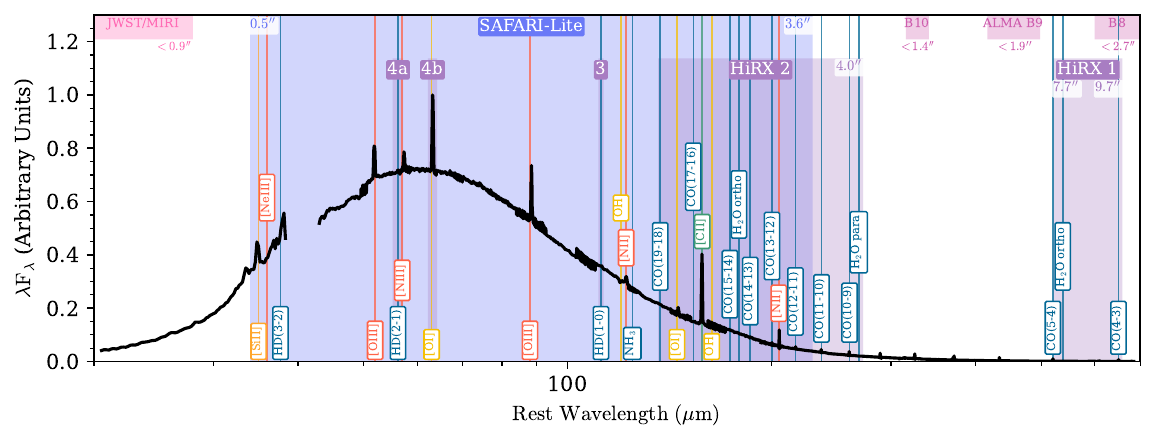}}
\caption{An example of the wavelength coverage of \saltus\ compared to a spectrum of the archetypal nearby starburst galaxy M\,82 (black; from Spitzer/IRS\citep{Kennicutt2003}, ISO/LWS\citep{Brauher2008}, and Herschel/SPIRE\citep{Kamenetzky2012}). The SAFARI-Lite simultaneous coverage is shown in blue and the HiRX bands are shown in purple. The coverage of JWST/MIRI (light pink) and the ALMA bands (dark pink) are plotted for reference. \change{The angular resolutions of each instrument at key wavelengths are shown in the corresponding colors.}  Key diagnostic lines are shown, tracing the ionized (red), neutral (yellow), molecular (blue), ionized/neutral (orange), and ionized/neutral/molecular (green) ISM phases; this list is not exhaustive. With \saltus, we will measure spectra of nearby galaxies at substantially higher spatial resolution, spectral resolution, and sensitivity compared to previous instruments, opening a fundamentally new window into the study of the ISM in the local Universe.} 
\label{fig:wlcoverage}
\end{figure*}

\change{The SAFARI-Lite instrument\citep{Roelfsema2024JATIS}} provides $R\approx300$ ($\Delta v\approx1000$~\kms) spectroscopy over the entire $34-230$\,\micron\ wavelength range simultaneously, covering the entire FIR at $z=0$ (Figure~\ref{fig:wlcoverage}). \change{To achieve this, the wavelength range is divided into four co-aligned bands. Light enters the instrument through six spatial pixels (arranged linearly on-sky) which are then dispersed onto 180-detector MKID arrays.} ISM heating and cooling is governed by forbidden transitions of atoms and molecules which occur in the FIR. The key line, and the brightest line in galaxies in the FIR, is the 158\micron\ transition of \CII. This line will be covered by both SAFARI-Lite and HiRX, enabling detection and characterization over a range of environments and targeted high spectral resolution kinematic/energetic studies. Other key lines covered by SAFARI-Lite include \SIII~34\micron, \SiII~35\micron, \OI~63,145\micron, \OIII~52,88\micron, \NII~122,205~\micron, and high-J CO transitions (Figure \ref{fig:wlcoverage}).

\change{The HiRX instrument\citep{daSilva2024JATIS}} provides sub-\kms\ spectral resolution in four tunable bands chosen to cover key spectral features for a broad range of science cases (Figure \ref{fig:wlcoverage}). Band 1 (521-659\micron) covers two key CO rotational transitions (J=5-4 and J=4-3) and a H$_2$O ortho transition. Band 2 ($136-273$\,\micron) covers key lines such as \NII~122,205\,\micron, \OI~145\,\micron, \CII~158\,\micron, several deuterated water (HDO) transitions, two prominent H$_2$O lines, and several high-J CO lines. Band 3 (112\,\micron) covers the ground state rotational transition of deuterated molecular hydrogen (HD 1-0). Band 4 ($56.2,63.2$~\micron) covers HD 2-1 and \OI~63\,\micron.

\subsection{Spatial Resolution}
\label{ssec:spatialres}

It is expected that \saltus\ would be diffraction limited down to 30\,\micron, achieving $0.5^{\prime\prime}-3.6^{\prime\prime}$ resolution over the wavelength range covered by SAFARI-Lite and $0.8^{\prime\prime}-9.7^{\prime\prime}$ resolution over the HiRX bands (Figure~\ref{fig:wlcoverage}). As a result, \saltus\ will offer transformative gains in physical resolution for the study of the Milky Way and nearby galaxies compared to the Herschel Space Observatory ($\approx16\times$ smaller beam area) or the Stratospheric Observatory for Infrared Astronomy (SOFIA; $\approx30\times$ smaller beam area). 
For example, \saltus\ will enable mapping of Milky Way sources, such as Orion BN/KL at $\approx200-1000$~AU spatial resolution, filling gaps between ISM and star and planetary system formation science questions. For galaxies within 20~Mpc (e.g., the PHANGS sample \citep{Leroy2021,Lee2023}), \saltus\ enables mapping of entire galaxies on $50-200$~pc scales, revealing the structure of the ISM and sites of feedback. \change{As indicated in Figure~\ref{fig:wlcoverage}, the spatial resolution achieved by SALTUS is extremely well-matched to JWST/MIRI ($\approx$0.9\arcsec\ at 28\micron) and ALMA ($<$1.4\arcsec\ at 345\micron; $<$2.7\arcsec\ at 650\micron).}

\subsection{Mapping Sensitivity}
\label{ssec:sensitivity}


\change{\saltus' large (14-m), passively cooled ($<40$~K) primary mirror offers transformative increases in sensitivity over previous instruments\citep{Chin2024JATIS}}. The sensitivity of SAFARI-Lite is similar or better than that of JWST/MIRI at the short wavelength end and ALMA at the long wavelength end. SAFARI-Lite would be more than two orders of magnitude more \change{sensitive} compared to previous comparable instruments, such as the Photodetector Array Camera (PACS) on Herschel\citep{Poglitsch2010} and Field-Imaging Far-Infrared Line Spectrometer (FIFI-LS) onboard SOFIA\citep{Fischer2018}. 

\subsection{Spectral Mapping Speed}
\label{ssec:mappingspeed}

Although \saltus\ has a large primary mirror, it is able to map a $\approx$5~arcmin$^2$ area without repointing by using a fine-steering motor. This map size is similar to those of single JWST/NIRCam and MIRI images. Crucially, however, these ``maps" from \saltus/SAFARI-Lite are three-dimensional, containing a spectrum covering $34-230$\,\micron\ at each pixel, and therefore contain significantly more information than single filter images from JWST. This mapping area is sufficient to capture Milky Way star-forming regions and many nearby galaxies in one or two pointings. Larger, more nearby targets, e.g., M\,82, NGC\,253, or Local Group galaxies, will need multiple pointings to observe the entire disk, but a single pointing is sufficient to study particular areas of interest (e.g., galactic centers, regions of high star formation, etc).




\section{Key Milky Way and Nearby Galaxies Science Goals for SALTUS}
\label{sec:keygoals}

SAFARI-Lite's high spatial resolution and simultaneous $34-230$\,\micron\ full spectral coverage, combined with HiRX's sub-\kms\ spectral resolution and efficient mapping capabilities using a scanning mirror, make \saltus\ the only observatory that can measure the key diagnostic lines of gas, and quantitatively address the roles of stellar feedback in the evolution of regions of massive star formation and of protostellar feedback during the earliest, deeply embedded phases of star formation, and determine how this depends on the cluster characteristics and core environment \change{(e.g., Figure~\ref{fig:lifecycle})}. \change{Here we present some sample science use cases which address the key DQs laid out in Section~\ref{sec:decadal}.}

\subsection{Understanding the Role of Feedback in Milky Way Star-forming Regions}
\label{sec:mw}

\begin{figure}[t]
\begin{center}
\begin{tabular}{c}
\hspace{-1mm}\includegraphics[width=0.9\textwidth]{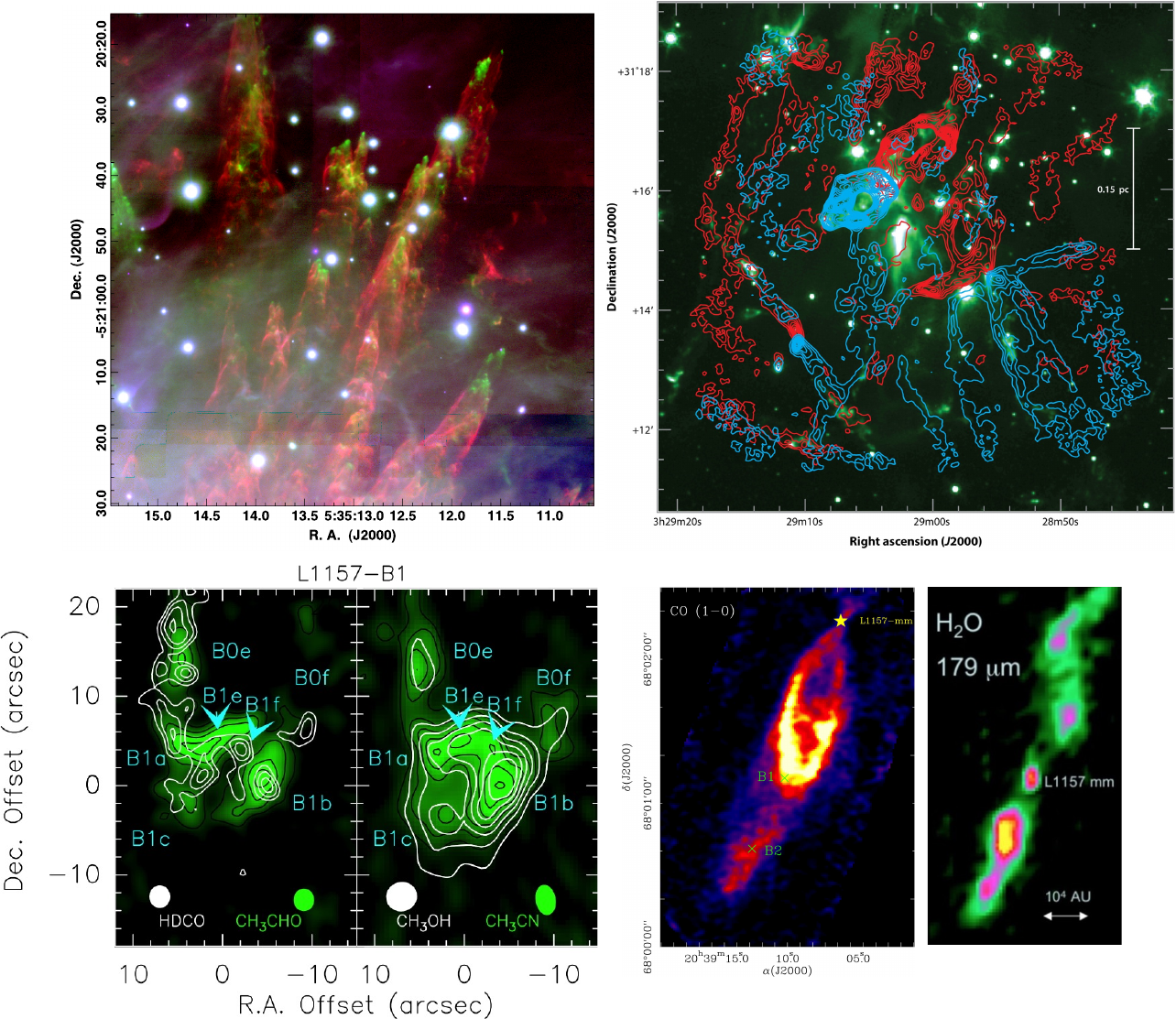}  
\\
\end{tabular}
\end{center}
\caption 
{\label{fig:D6}
{Top row: (Left) Part of the Northern (peak 1) outflow lobe in the BN/KL region as traced in the K band (blue), 1.644$\mu$m [FeII] (green), and 2.122$\mu$m H$_2$ (red), revealing well defined H$_2$ emission with [FeII] fingertips\citep{Bally2015}. (Right) Outflows driven by the young stellar cluster in the NGC\,1333 region. Some 22 outflow lobes are evident in the 4.5$\mu$m IRAC image (green) and the red and blue shifted CO J=1-0 emission in this 0.5~pc field\citep{Plunkett2013}. Bottom row: The bipolar outflow associated with the proto star L1157mm. (Right) H$_2$O emission in the ground-state ortho line at 179$\mu$m obtained by {\em Herschel}/PACS at a resolution of 13\arcsec, revealing unresolved emission clumps\citep{Nisini2010}. (Middle) A blow up of the blue shifted, southern lobe in the CO J=1-0 transition at a resolution of 3\arcsec. Two bright emission structures, B1 \& B2, are indicated\citep{Busquet2014}. (Left) Interferometric studies of molecular emission in the B1 clump break this structure up into multiple components\citep{Codella2015}. Green contours trace acetaldehyde and methyl cyanide, respectively, while white is deuterated formaldehyde and methanol. The clear chemical differentiation of these complex organic molecules reflects the effects of the interplay of shock sputtering of icy grains and subsequent chemistry in the warm gas. Beam sizes ($\sim$3\arcsec) are indicated and comparable to SAFARI-{\em Lite} at the longest wavelengths.}} 
\end{figure} 

A key \saltus\ science goal aims to understand the role of feedback in star forming regions, both in nearby Milky Way regions as well as in nearby spiral galaxies. It quantifies the kinetic energy and momentum input during the earliest, deeply embedded phases of star formation, and addresses how this depends on the cluster characteristics and core environment \change{(DQs 1a, 3b, 4b)}. 

Orion~BN/KL and NGC\,1333 are the two prototypical Galactic sources of star formation to be studied that demonstrate SALTUS’s unique capabilities to study protostellar outflows to the general community. The combination of high spatial resolution, simultaneous full spectral coverage, and mapping efficiency make SALTUS the only observatory that can measure the key diagnostic lines of interstellar gas and quantitatively address the role of protostellar feedback during the earliest, deeply embedded phases of star formation and how this depends on the cluster characteristics and core environment \change{(DQ 1a and 3b)}. ALMA can measure low J CO sub-millimeter transitions at high spatial resolution, but the high J transitions that are key to measuring the temperature, density, and column density of warm, dense shocked gas occur at far-IR wavelengths, are only accessible with SALTUS \change{(DQ 2a)}. JWST/MIRI can measure the pure rotational transitions of H$_2$ in the mid-IR that are good diagnostics of the shock characteristics as well. However, the small foot-print of MIRI spectroscopy($<$7.7\arcsec) precludes efficient mapping of regions of extended emission associated with protostellar outflows. Another probe of these outflows can be achieved through spectral-mapping in key diagnostic lines of J and C shocks ([OI], H$_2$O, high J CO) using SAFARI-Lite and HiRX, which probes a range of cluster characteristics and molecular core properties\citep{Hollenbach1989,Kaufman1996,Flower2013}.

The BN/KL region in Orion is the nearest site of massive star formation, presenting an excellent opportunity to study in detail the interaction of massive protostars with the surrounding molecular core \change{(DQs 1a and 3b)}. Scaling the Herschel/PACS fluxes for peak 1\citep{Goicoechea2015}, assuming homogeneous emission, results in expected line fluxes in the 1-3\arcsec\ SALTUS beam of $3\times10^{-18}$ to $2\times10^{-15}$~W~m$^{-2}$ for selected $^{12}$CO, H$_2$O and OH lines and the [OI]~63$\mu$m fine-structure line. In $^{13}$CO, SALTUS can be expected to detect lines as high as J=30-29 (E$_u\sim2500$~K). SAFARI-Lite can map the 5~arcmin$^2$ of peaks 1 and 2 in Orion down to a limiting line flux of $1-3\times10^{-18}$~W~m$^{-2}$ in 5~hours. This data can be analyzed to determine column density, preshock density, and shock velocity\citep{Goicoechea2015} on a $1-3$\arcsec\ scale and quantify the disruption of the Orion Molecular Cloud (OMC) 1 core by this explosion and its effect on the surrounding star cluster.

NGC\,1333 is the most active region of low-mass star formation in the Perseus molecular cloud, containing a large-membership ($>$100) young clusters, and is considered the prototypical, nearby (135~pc) cluster-forming region\citep{Walawender2008,Padoan2009}. NGC\,1333 contains many molecular outflows (Figure~\ref{fig:D6})\citep{Bally1996,Sandell2001,Davis2008} that appear capable of maintaining turbulence and therefore limiting the rate of star formation\citep{Arce2010,Nakamura2012,Bally2014}. Typical Herschel/PACS high-J CO and H$_2$O line fluxes of the shock emission associated with the protostellar outflows in NGC\,1333 are \change{$\sim10^{-16}$~W~m$^{-2}$}\citep{Herczeg2012} and translate into average line fluxes of $6\times10^{-18}$~W~m$^{-2}$ in the SALTUS beam. A spectral-spatial study with SAFARI-Lite can map a 1~arcmin$^2$ and detect these lines at S/N ratio 30. Spitzer/IRS H$_2$ studies\citep{Maret2009} imply that the shocked gas comprises in total some 5~arcmin$^2$. A fully sampled $34-230$~\micron\ spectral-spatial map of the shocked gas in NGC\,1333 takes 5 hours with SAFARI-Lite. These data can be analyzed analogously to the Herschel data using existing shock models and provide on a $\sim$3\arcsec\ scale size the shock characteristics (preshock density, shock temperature, shock velocity) and quantify the energy and momentum provided by the protostellar outflows for the NGC\,1333 cluster \change{(DQs 1a and 3b)}. Combination of these data with ground-based interferometric maps in complex organic molecules can then address the influence of shocks on the organic inventory of star forming cores \change{(DQs 2b and 3a)}.


\subsection{Understanding the Role of Feedback in Nearby Galaxies}
\label{sec:nearbygal}

\begin{figure}[t]
\begin{center}
\begin{tabular}{c}
\hspace{-1mm}\includegraphics[width=\textwidth]{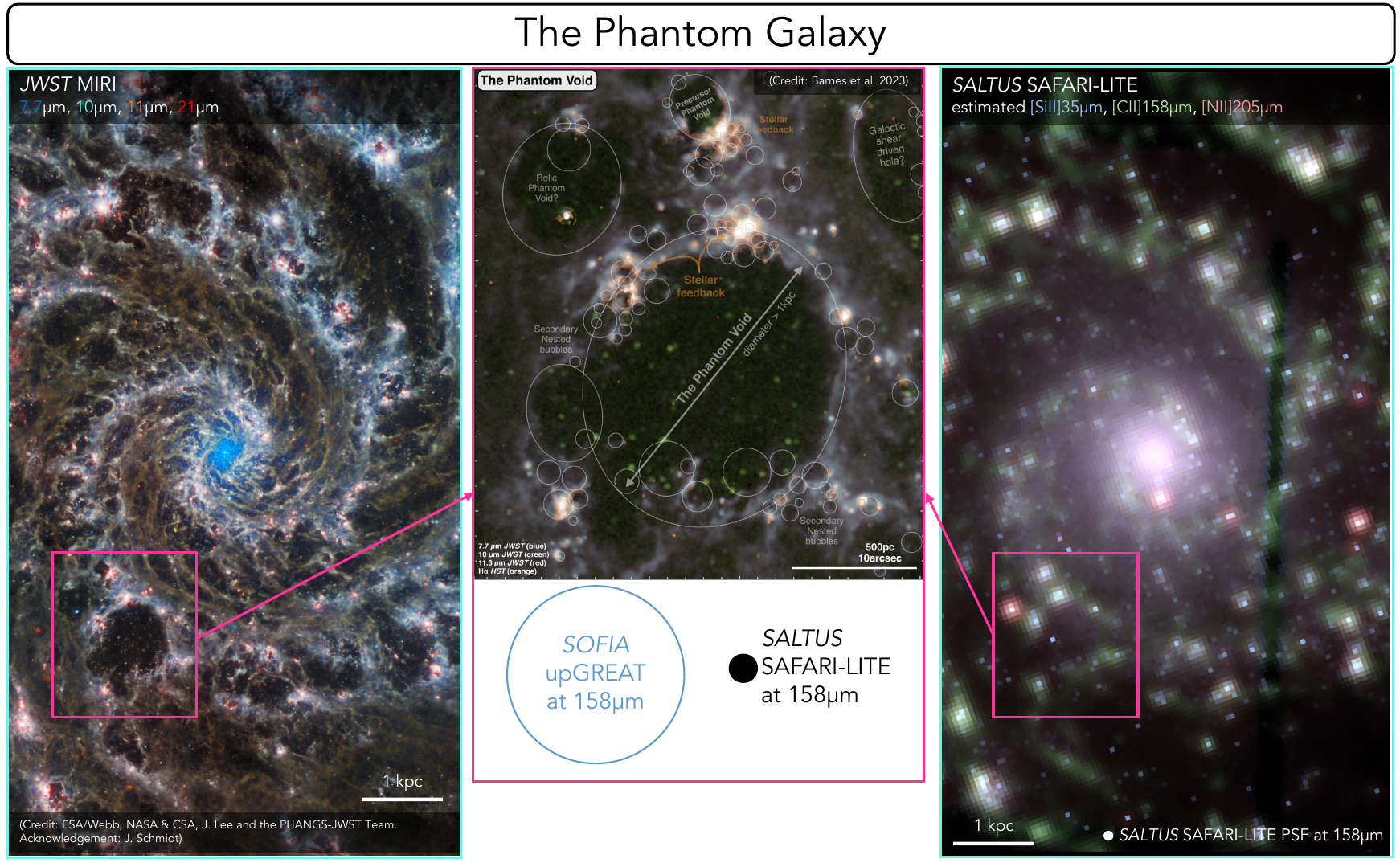}  
\\
\end{tabular}
\end{center}
\caption 
{\label{fig:FO11}
{JWST has revealed new details in the structure of the interstellar medium (ISM) by observing the dust emission of NGC\,628. (Left) The concerted action of stellar winds and supernovae create superbubbles, typically $<$500~pc across, that drive the evolution of the galactic ecosystem (Image credit: ESA/Webb, NASA, \& CSA, J. Lee and the PHANGS-JWST Team). (Center) \saltus\ will complement these JWST studies, extending high resolution observations of the ISM into the FIR. (Image credit: \citep{Barnes2023}). (Right) As the simulated SAFARI-Lite image illustrates, \saltus\ will simultaneously measure key diagnostic atomic fine-structure lines like the three simulated here, resolving these bubbles and quantifying the physical conditions and energetics of this feedback.}} 
\end{figure} 

The second part of this science case bridges ``ground-truth" observations of feedback on the scale of individual Milky Way star-forming regions to the spatially unresolved view of galaxy evolution throughout the distant universe \change{(DQ 4a)}. 
Energetic feedback from star formation processes, the lives and deaths of stars, and nuclear black holes (or active galactic nuclei, AGN) are critical factors in the growth and evolution of galaxies and the gas that surrounds them. The effects of feedback are multiscale, ranging from individual stars and clusters to galaxy-scale outflows extending several kiloparsecs into the circumgalactic medium. However, the injection sites of feedback are small-scale, originating from individual supernovae, star clusters, and near the AGN itself (Figure~\ref{fig:FO11}).

Spectral line strengths and ratios (e.g., \SiII~35\micron, \OI~63, 145\micron, \OIII~52, 88\micron, \NII~122, 205\micron) are used as diagnostics of radiation field, temperature, and/or density to characterize shocks and feed-back in the ISM, especially when coupled with modeling codes such as {\tt CLOUDY}\citep{Tarantino2021}. The \CII\ line at 158\micron\ is a tracer of the neutral ISM, photodissociation regions, and star formation\citep{Crawford1985,Goldsmith2012,Herrera-Camus2018}. \CII\ and \OI~63$\mu$m are major cooling channels for the ISM. The \OIII\ and \NII\ lines are tracers of ionized gas in \HII\ regions and the diffuse ISM, respectively. Since FIR lines largely govern the bulk heating and cooling of the ISM, they are key to understanding energy balance in the ISM, especially in regions of active feedback. 

Figure~\ref{fig:FO11} gives an example in a nearby, star-forming galaxy M\,74 (The Phantom Galaxy or NGC\,628). While new observations from JWST have resolved the galaxy and sites of stellar feedback in stunning detail (Figure~\ref{fig:FO11})\citep{Lee2023,Watkins2023,Barnes2023}, the physics of the photodissociation regions surrounding these supernova bubbles is unknown without FIR diagnostics. To understand how ISM heating and cooling vary under such conditions requires high spatial resolution observations of key FIR diagnostics \change{(DQ 1a and 4b)}. SAFARI-Lite enables such detailed studies of the effects of stellar and AGN feedback across a suite of nearby galaxies. The capabilities are highlighted in Figure~\ref{fig:FO11} (right), showing the anticipated \CII\ and \NII~205$\mu$m emission from M\,74 at SAFARI-Lite resolution. The \CII\ emission is estimated from the continuum-subtracted 3.3$\mu$m PAH feature\citep{Sandstrom2023}, and the \NII\ emission is based on H$\alpha$\citep{Elmegreen2006}. A 100$\sigma$ detection of the \CII~158$\mu$m line (to enable robust detection of weaker lines and continuum) over the full 9\arcmin$\times$9\arcmin\ extent of M\,74 requires 12 hours.

HiRX enables high spectral resolution (R$>10^6$, $\Delta$v$<$1~km/s) observations of strong emission lines (e.g., \CII~158$\mu$m, \NII~205$\mu$m, \OI~145$\mu$m, H$_2$O, several high-J CO lines; Figure~\ref{fig:wlcoverage}). The precision kinematics yield robust measurements of energy and momentum injection from feedback and shocks on scales ranging from star clusters to large-scale outflows\citep{Levy2021,Levy2022} \change{(DQs 1a, 3a, 4b)}. In M\,74 specifically (Figure~\ref{fig:FO11}), the expansion of the ``Phantom Void" is $15-50$~\kms. HiRX will be the only instrument capable of spatially and spectrally resolving feedback-driven bubbles, which are expected to be commonplace in star-forming galaxies\citep{Watkins2023,Barnes2023}.

\subsection{Example Community Science: Stellar Feedback in Regions of Massive Star Formation}

\begin{figure}[t]
\begin{center}
\begin{tabular}{c}
\hspace{-1mm}\includegraphics[width=\textwidth]{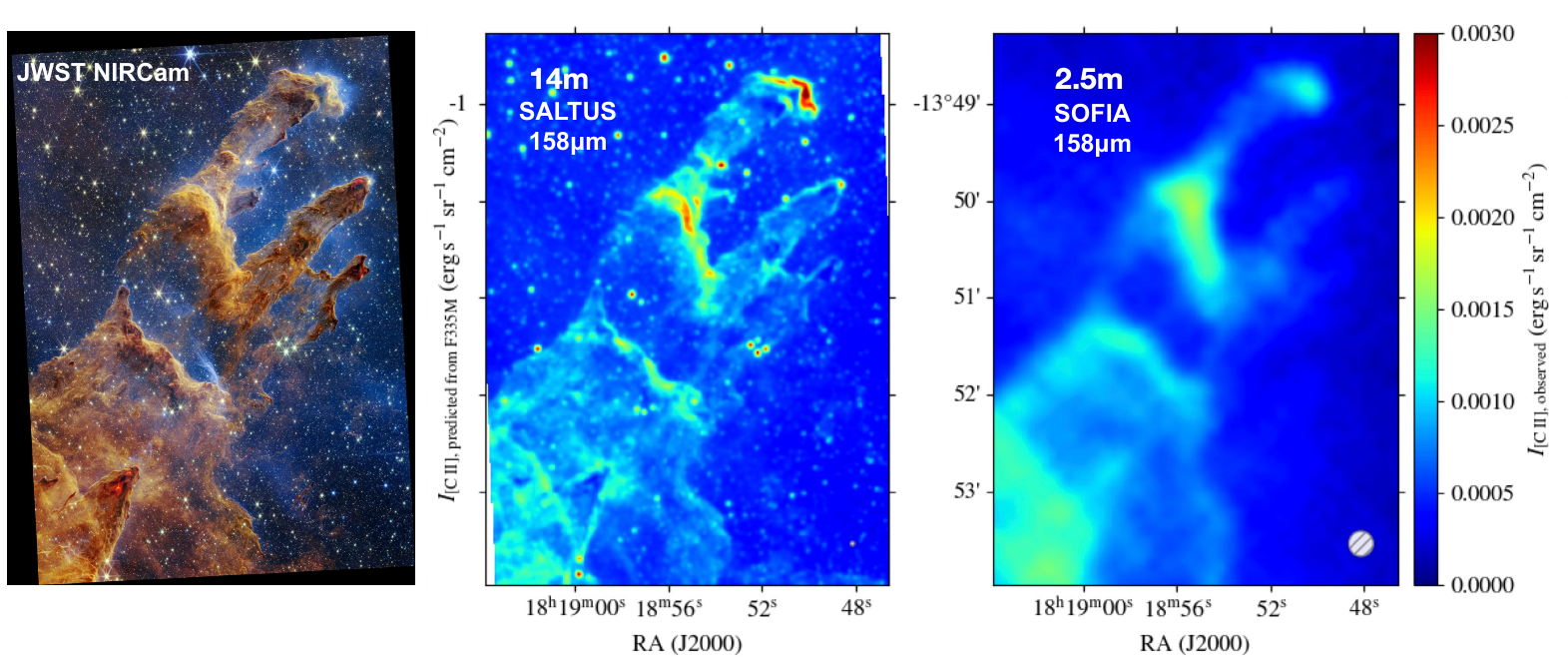}  
\\
\end{tabular}
\end{center}
\caption 
{\label{fig:D3}
{Simulated SALTUS image at 2.5\arcsec\ angular resolution (middle) of the [CII] 158$\mu$m emission in NGC\,6611 (Pillars of Creation) is similar to the JWST NIRCam image (left) and compared to the SOFIA-created map\citep{Karim2023} (right). SAFARI-{\em Lite} can map this 10~arcmin$^2$ region in 10 hours and simultaneously provide maps in all diagnostic lines and photo-dissociation regions (PDRs) and HII regions in our galaxy and the local group probing the physical environment produced by radiation feedback of massive stars and its link to stellar clusters and its molecular core.}} 
\end{figure} 

This topic aims to understand the role of stellar feedback in regions of massive star formation: How much kinetic energy is contributed to the ISM, how does this depend on the characteristics of the region, and, connected to this, how does the surrounding medium react \change{(DQ 1a, 2a, 3b, 4b)}? This objective is addressed by spectral-spatial mapping of regions of massive star formation in the dominant far-IR cooling lines of PDR gas using both the SAFARI-Lite and HiRX instruments (c.f., Pillars of Creation in Figure~\ref{fig:D3}). The sample consists of 25 regions with typical sizes of 25~arcmin$^2$, covering a range in stellar cluster properties (single stars to small clusters to superstar clusters), GLIMPSE bubble morphology (ring-like, bi-or multi-polar, complex), environment (isolated star forming region, galactic mini starburst, nuclear starburst), and cluster age (0.1 to 5 Myr).

SAFARI-Lite will provide full spectra over the 34-230$\mu$m range, containing the key diagnostic atomic fine-structure transitions. SOFIA/upGREAT studies of the [CII]~158$\mu$m emission of regions of massive star formation have demonstrated the feasibility of this technique\citep{Luisi2021,Tiwari2021}. Together these two data sets quantify the kinetic energy of the expanding shell as well as the thermal, turbulent, and radiation pressures on the shell that can be directly compared to radiative and mechanical energy inputs of the stellar cluster.

\subsection{Example Community Science: Putting Early-Universe Feedback in Context with Local Analogs of the First Galaxies}

The first year of JWST observations revealed that early generations of galaxies were typically much lower in mass and significantly less metal-enriched compared to the massive star-forming spiral galaxies that dominate star formation today. Consequently, their UV radiation fields were harder\citep{Endsley2023}, which impacted the gas ionization state and cooling rate. The shallow gravitational potential wells also enhanced the effects of supernova-driven feedback since single supernovae could eject up to $\sim$95\% of the heavy elements formed during the star’s lifetime into the circumgalactic medium\citep{Peeples2014,McQuinn2015}. While ALMA now allows the key far-IR fine structure diagnostic lines (e.g., [CII], [OIII]) to be detected at high redshifts, these lines still take several hours to detect even in the most massive and UV-luminous reionization-era galaxies\citep{Bouwens2022} and are generally out of reach for the more typical galaxies thought to drive cosmic reionization.

Thanks to community efforts, many nearby galaxy populations have been identified bearing similarities to UV-bright reionization-era galaxies in terms of their mass, metallicity, and harsh UV radiation fields. Though rare locally and largely unknown during the time of Herschel, these low-redshift galaxies are plausible analogs to early star-forming galaxies. Such local low-metallicity dwarf galaxies are natural extensions to the SALTUS low-redshift GTO efforts to measure and resolve the injection of feedback energy on small spatial scales far beyond the capabilities of Herschel\citep{Madden2013} or a 1m-class FIR mission. With the same observables as the GTO program of star-forming spirals and starbursts, and the most commonly detected lines found with ALMA at high redshifts, SALTUS enables $\sim$200~pc-scale maps of the FIR fine structure lines and underlying dust continuum in low-redshift analogs of the first generation of galaxies \change{(DQs 1a, 3b, 4a, 4b)}.

\subsection{Example Community Science: Spectral Line Survey}

The benefits of performing high-frequency spectral line surveys were recognized and exploited by Herschel HIFI, which resulted in the first reported detections of SH$^+$, HCl$^+$, H$_2$O$^+$, and H$_2$CL$^+$ \citep{Benz2010,DeLuca2012,Ossenkopf2010}. The formation of these small hydrides typically represents the first step in gas phase chemistry routes toward molecular complexity in space \change{(DQ 2b)}. However, the spectral surveys with HIFI were severely limited by sensitivity. SALTUS’s increased aperture results in a 16$\times$ increased sensitivity for these unresolved sources and can be expected to lead to an enhanced line density and discovery space \change{(DQ 4b)}. Moreover, the wavelength coverage of \saltus\ bridges the gap between spectral line surveys to be carried out with JWST/MIRI and ALMA\citep{Martin2021}.

\section{Conclusions} \label{conclusions}
Unlocking the details of ISM heating and cooling, feedback, and star-formation as a function of environment are keys to understanding our {\em Cosmic Ecosystem}\citep{astro2020} and rely on measurements that can only be made in the FIR. While any future FIR facility will offer transformative gains over previous observatories at this wavelength, only \saltus\ will enable the spatially-resolved studies in nearby galaxies to probe a wider range of conditions and to inform numerical simulations. Priority Milky Way measurements include measuring PDRs and feedback in prototypical high- and low-mass star forming regions at unprecedented resolution and sensitivity, bridging the gaps between studies of star and planet formation and the ISM. In nearby galaxies, the priority is to measure PDRs and feedback in well-characterized prototypes (e.g., M\,74) across a range of environments. \saltus' transformative increase in spatial resolution, spectral coverage, and sensitivity at these wavelengths ensure that it will be responsive to the needs of and bridge gaps between the Galactic and nearby extragalactic communities of the 2030s. No other FIR facility will provide such seamless continuity in our understanding of the ISM after a decade of JWST observations and ALMA/ALMA-2030. 
 
\subsection*{Disclosures}
The authors have no relevant financial interests in the manuscript nor do they have any conflicts of interest to disclose.

\subsection*{Code and Data Availability}
No code or data were used in the preparation of this manuscript.

\subsection* {Acknowledgments}
R.C.L. acknowledges support for this work provided by a National Science Foundation (NSF) Astronomy and Astrophysics Postdoctoral Fellowship under award AST-2102625.
This work has made use of NASA's Astrophysics Data System Bibliographic Services.
This work has made use of the NASA/IPAC Extragalactic Database (NED), which is funded by NASA and operated by the California Institute of Technology.

\bibliographystyle{spiejour}   


\vspace{2ex}\noindent\textbf{Rebecca C. Levy} is a NSF Astronomy \& Astrophysics Postdoctoral Fellow at the University of Arizona. She received her BS in Astronomy and Physics from the University of Arizona in 2015, and her MS and PhD degrees in Astronomy from the University of Maryland, College Park in 2017 and 2021, respectively. She has authored more than 40 refereed journal papers. Her research interests include studying extreme star formation, star clusters, stellar feedback, and the interstellar medium in nearby galaxies using multiwavelength ground- and space-based observations. 

\vspace{2ex}\noindent\textbf{Alexander Tielens}  is a professor of astronomy in the Astronomy Department of the University of Maryland, College Park. He got his masters and PhD in astronomy from Leiden University in 1982. He has authored over 500 papers in refereed journals and written two textbooks on the interstellar medium. His scientific interests center on the physics and chemistry of the interstellar medium, in particular in regions of star and planet formation, including the characteristics and origin and evolution of interstellar polycyclic aromatic hydrocarbon molecules and dust grains. He has published extensively on the radiative and mechanical feedback of massive stars on their environment; especially, the observational and theoretical characteristics of photodissociation regions.

\vspace{2ex}\noindent\textbf{Justin Spilker} \change{is an assistant professor in the Department of Physics and Astronomy and the Mitchell Institute of Texas A\&M University. He is interested in the quenching of galaxies - the processes that prevent galaxies from forming new stars and to keep them from forming stars over long timescales. He uses radio/submillimeter interferometers like ALMA and the VLA for these studies. He has studied a sample of very dusty, highly starforming galaxies detected by the South Pole Telescope that are magnified by a foreground galaxy through gravitational lensing. By virtue of their selection, this sample tends to lie at higher redshift than other samples observed with either Herschel or SCUBA/SCUBA-2. He has authored about 180 publications.}

\vspace{2ex}\noindent\textbf{Daniel Marrone} \change{is a professor and experimental astrophysicist in the Department of Astronomy/Steward Observatory at the University of Arizona. He received his PhD at Harvard in 2006. His research addresses galaxies and cosmology and fundamental physics through a variety of observational tools. He is particularly interested in galaxy clusters and their cosmological applications, galaxy formation in the early universe, and the physics of the supermassive black hole in Sagittarius A*. He has developed new instruments, primarily at centimeter to submillimeter wavelengths. He has authored about 190 publications.}

\vspace{2ex}\noindent\textbf{Desika Narayanan} is an associate professor of astronomy at the University of Florida, and an affiliate of the Cosmic Dawn Center in Copenhagen.  He mainly studies the physics of the interstellar medium in galaxies near and far via large scale numerical simulations of galaxy evolution.  He earned his BS from the University of Florida in 2003, his PhD from the University of Arizona in 2007, and was the CfA fellow at the Harvard-Smithsonian Center for Astrophysics, as well as the Bok Fellow at the University of Arizona.  He was a professor at Haverford College from 2014-2017, prior to returning to the University of Florida.

\vspace{2ex}\noindent\textbf{Christopher K. Walker} \change{is a professor of Astronomy, Optical Sciences, Electrical \& Computer Engineering, Aerospace \& Mechanical Engineering, and Applied Mathematics at the University of Arizona. He received his M.S.E.E. from Clemson University (1980), M.S.E.E. from Ohio State University (1981), and Ph.D. in Astronomy from the University of Arizona (1988). He has worked at TRW Aerospace and the Jet Propulsion Laboratory, was a Millikan Fellow in Physics at Caltech, and has been a faculty member at the University of Arizona since 1991. He has made many contributions to advance the field of terahertz astronomy. He has supervised sixteen Ph.D. students, led numerous NASA and NSF projects, authored/coauthored 130+ papers, and published two textbooks: ``Terahertz Astronomy" and ``Investigating Life in the Universe". He is the principal investigator of the SALTUS mission concept.}

\vspace{1ex}
\noindent Biographies and photographs of the other authors are not available.

\listoffigures
\listoftables

\end{spacing}
\end{document}